\documentclass[twocolumn,english,showpacs]{revtex4}
\usepackage{babel,amsmath,amssymb,dcolumn}
\usepackage[dvips]{graphics}

\begin{document}

\title{Particle motion around magnetized black holes: Preston-Poisson space-time.}

\author{R.A. Konoplya}
\email{konoplya@fma.if.usp.br}
\affiliation{Instituto de F\'{\i}sica, Universidade de S\~{a}o Paulo \\
C.P. 66318, 05315-970, S\~{a}o Paulo-SP, Brazil}


\begin{abstract}
We analyze motion of massless and massive particles around black
holes immersed in an asymptotically uniform magnetic field and
surrounded by some mechanical structure, which provides the
magnetic field. The space-time is described by Preston-Poisson
metric, which is the generalization of the well-known Ernst metric
with a new parameter, tidal force, characterizing the surrounding
structure. The Hamilton-Jacobi equations allow separation of
variables in the equatorial plane. The presence of tidal force
from surroundings considerably changes parameters of the test
particle motion: it increases the radius of circular orbits of
particles, increases the binding energy of massive particles going
from a given circular orbits to the innermost stable orbit near
black hole. In addition, it increases the distance of minimal
approach, time delay and bending angle for a ray of light
propagating near black hole.
\end{abstract}

\maketitle




\section{1. Introduction.}

Black holes in the centres of galaxies are immersed in a strong
magnetic field due to charged matter surrounding them. The strong
magnetic field in the centre of galaxies is stipulated by toroidal
currents around galactic black holes \cite{AGN}. Therefore an exact solution
of Einstein-Maxwell equations describing a black hole immersed in
an asymptotically uniform magnetic field, known as Ernst solution
\cite{Ernst}, was of considerable interest \cite{MBH}.
The light and particle motion around Ernst-Schwarzschild black
hole was analyzed in a few papers \cite{Galtsov-Ernst},
\cite{Esteban}. In particular, in \cite{Galtsov-Ernst} it was
shown that Hamilton-Jacoby equation allow separation of variables
in the equatorial plane, where the motion of neutral and charged
particles were analyzed. In \cite{Esteban} the motion of neutral
particles were considered for a more general situation of
electromagnetized Kerr background. There it was shown that the
release of binding energy is considerably increased because of
presence of electromagnetic field, and the binding energy for
circular orbits was calculated. Yet, in a more realistic
situation, the strong magnetic field in the central region near
black hole is created by some surrounding matter, such as
accretion disk or an active galactic nuclei.  This surrounding
structure exerts strong gravitational tidal force on particles
moving near black holes, so that magnetic influence of the
structure might be even much smaller than its gravitational
influence. Therefore a more physical situation should include into
consideration, the corrections to the black hole metric due-to
that structure. Fortunately, recently Preston and Poisson
\cite{Preston-Poisson1} have found such a corrected metric.
This is the solution to the perturbative Einstein-Maxwell
equations depending on three parameters: the black hole mass $M$,
magnetic field $B$, and a new parameter $K$, which characterize
the above surrounding structure. The solution is very accurate for
$r^{2} B^{2} \ll $, $M/a \ll 1$,  and $r{^2} K \ll 1$, where $r$
is the distance form the black hole, $a$ is the length scale of
the mechanical structure. Indeed, comparison with the exact Ernst
solution shows that next order corrections are of order $B^{4}$,
and are very small.

In the present paper we generalize analysis of works
\cite{Esteban} and \cite{Galtsov-Ernst} and study motion of test
particles near black holes immersed in asymptotically uniform
magnetic field and some gravitating surrounding structure, which
provides the magnetic field. The paper is organized as follows: In
Sec. II we reduce the Preston-Poisson metric to the Ernst-like
form, by going over to a new coordinates. Then in Sec. III we
consider the Hamilton-Jacobi equation in the equatorial plane, and
use it for analysis of massless particles. There the lens effects
for Preston-Poisson metric is considered. Motion of massive
particles is described in Sec. IV, where the binding energy for
particles on circular orbits are calculated.

\section{Preston-Poisson metric}

Following \cite{Preston-Poisson1}, let us consider a model
consisting of a non-rotating black hole immersed in a uniform
magnetic field, and a large mechanical structure, such as a giant
solenoid, producing the magnetic field of strength $B$. The
structure has a mass $M'$ and its linear extension is $\sim a$. In
order to have magnetic field which is uniform for $r \gg M$, one
chooses
\begin{equation}
A^{\alpha} \frac{1}{2} B \phi{^\alpha},
\end{equation}
where $\phi{^\alpha}$ is the rotational Killing vector of the
unperturbed Schwarzschild metric.

The metric which describes the space-time of the above model,
written in light-cone gauge, is \cite{Preston-Poisson1}
\begin{equation}
d s^{2} = - g_{v v} d v^{2} + 2 d v d r + g_{v \theta} d v d \theta +
g_{\theta \theta} d \theta^{2} + g_{\phi \phi} d \phi^{2},
\end{equation}
where
$$ g_{v v} = -(1 - (2 M/r)) - \frac{1}{9} B^{2}r(3 r - 8 M)-
$$
\begin{equation}
\left(\frac{1}{9}B^{2}(3 r^{2} -14 M r + 18 M^{2})+ K (r - 2 M)^{2}\right)(3
\cos^{2} \theta -1),
\end{equation}
\begin{equation}
g_{v \theta} = \left(\frac{2}{3} B^{2}(r- 3 M)- 2 K\right)r^{2}
\sin\theta \cos\theta
\end{equation}
\begin{equation}
g_{\theta \theta} = r^{2} + \left(- \frac{1}{3} B^{2} r^{2} +
B^{2} M^{2} + K (r^{2}- 2 M^{2})\right) r^{2}\sin^{2}\theta
\end{equation}
$$ g_{\phi \phi} = r^{2} \sin^{2}\theta + $$
\begin{equation}
B^{2} r^{2} \left( \left(-
\frac{1}{3} r^{2} - M^{2}\right)- K  (r^{2}- 2 M^{2})) r^{2}
\right) \sin^{4}\theta
\end{equation}
The above metric is accurate through order ($B^{2}$, $K$),
whenever $r^{2} B^{2} \ll $ and $r{^2} K \ll 1$. The new parameter
$K$ characterizing the mechanical structure, containing the black
hole, can be interpreted as a tidal gravity or Weyl-curvature
when $r \gg M$. Thus we have a three-parameter solution.

From now on, we shall consider motion in the equatorial plane
$\theta =\pi/2$. Therefore, we shall put $\theta =\pi/2$ in
formulas (1)-(5). Then, let us make the following coordinate
transformations:
\begin{equation}
v= t + \overline{r} + 2 M \ln|(\overline{r}/2 M)-1|,
\end{equation}
$$ r = \overline{r} (1+ (1/6) B^{2} r^{2} +
 (1/3) (K - (1/2)B^{2}) (\overline{r}- 2M)
\overline{r}^{2}+ $$
\begin{equation}
 O(B^{4},K^{2}).
\end{equation}
These transformations are generalization of transformations
(3.60-3.61) of \cite{Preston-Poisson1} in the equatorial plane and
they cast the metric (1) into the diagonal form
\begin{equation}
d s^{2} = g_{t t} d t^{2} + g_{\overline{r} \overline{r}} d
\overline{r}^{2} + g_{\theta \theta} d \theta^{2} + g_{\phi \phi} d \phi^{2},
\end{equation}
where the metric components  (neglecting orders $O(B^{4},K^{2})$
and higher) are
\begin{equation}
g_{\overline{r} \overline{r}} =(1-2 M/\overline{r})^{-1} +
\frac{(4 M - 3 \overline{r}) \overline{r}^{2}}{6 M -3
\overline{r}}K - \frac{2 M \overline{r}^{2}}{6 M -3
\overline{r}} B^{2},
\end{equation}
$$ g_{t t} = (1-2 M/\overline{r}) + (1/3) (- 8 M^{2} + 10 M
\overline{r} - 3
\overline{r}^{2}) K + $$
\begin{equation}
 (2/3) M (2 M -\overline{r}) B^{2},
\end{equation}
$$ g_{\theta \theta} =\overline{r}^{2} + (1/3) \overline{r}^{2} (- 6
M^{2} - 4 M
\overline{r} + 5 \overline{r}^{2}) K +  $$
\begin{equation}
(1/3) \overline{r}^{2} (3 M - \overline{r})  (M +
\overline{r}) B^{2},
\end{equation}
$$ g_{\phi \phi} =\overline{r}^{2} - (1/3) \overline{r}^{2} (- 6
M^{2} + 4 M
\overline{r} + 5 \overline{r}^{2}) K - $$
\begin{equation}
 (1/3) \overline{r}^{2} (3 M^{2} - 2 M \overline{r} + \overline{r}^{2})
B^{2}.
\end{equation}
This form of the metric does not have non-diagonal components in
the equatorial plane and is much simpler for consideration of
motion of test particles. To be exact, non-diogonal components are
of order ($B^{4}$, $K^{2}$) and higher, and therefore can be
safely neglected.

\section{Motion of massless particles}

From now and on we shall write $r$ instead of $\overline{r}$. The
four-momentum is
\begin{equation}
p_{\mu} = g_{\mu \nu} \frac{d x^{\mu}}{d s},
\end{equation}
where $s$ is an invariant affine parameter. The Hamiltonian has
the form
\begin{equation} H=\frac{1}{2} g^{\mu \nu} p_{\mu}
p_{\nu}.
\end{equation}
The action can be represented in the form:
\begin{equation}
S = -\mu s -E t + L \phi + S_{r}(r) + S_{\theta}(\theta),
\end{equation}
where $E$ and $L$ are the particle`s energy and angular momentum
respectively.

Then, the Hamilton-Jacobi equations for null geodesics read
\begin{equation}
\frac{1}{2} g^{\mu \nu} \frac{\partial S}{\partial x^{\mu}} \frac{\partial S}{\partial
x^{\nu}} = - \frac{ \partial S} {\partial s} = \mu^{2}.
\end{equation}

It is evident that the equations of motions allow separation of
variables in the equatorial plane $\theta = \pi/2$. The first
integrals of motion are
\begin{equation}
\mu g_{r r} \frac{d r}{d s} = \pm \sqrt{-\frac{g_{rr}}{g_{tt}} (E^{2}-
U_{eff}^{2})},
\end{equation}
\begin{equation}
p_{\phi} = \mu g_{\phi \phi} \frac{d \phi}{d s} = L,
\end{equation}
\begin{equation}
p_{t} = \mu g_{t t} \frac{d t}{d s} = -E,
\end{equation}
\begin{equation}
U_{eff}^{2} = - g_{tt} \mu^{2}\left(1+ \frac{L^{2}}{\mu^{2}
g_{\phi
\phi}}\right)
\end{equation}

The trajectory and propagation equations take the form
\begin{equation}
\left(\frac{d r}{d t}\right)^{2} = -\frac{g_{tt}}{g_{rr}}
\left(1 + \frac{g_{tt}}{g_{\phi \phi}} \left(\frac{p_{\phi}}{p_{t}}\right)^{2}
-g_{t t} \frac{m^{2}}{p_{t}^{2}}\right),
\end{equation}
\begin{equation}
\left(\frac{d r}{d \phi}\right)^{2} = -\frac{g_{\phi \phi}}{g_{r r}} \left(1 + \frac{g_{\phi \phi}}{g_{t t}}
\left(\frac{p_{t}}{p_{\phi}}\right)^{2} -g_{\phi \phi } \frac{m^{2}}{p_{\phi}^{2}}\right).
\end{equation}

For massless particles, from the above equations (22), (23), one can see
that propagation and trajectory equations contain only ratio  $b =
L/E$, which is called the impact parameter. The qualitative
description of the motion of massless particles can be made by
considering the effective potential of the motion (21, where $\mu
=0$. The equation for radii of circular orbits can be found from
the condition $dU_{eff}/dr =0$:
\begin{equation}
g_{t t} g_{\phi \phi},_{r} = g_{\phi \phi} g_{t t},_{r}.
\end{equation}
This gives the algebraic equation for $r$:

\begin{widetext}

\begin{equation}
-6(B^2-2 K)M^3-2r + 10(B^2-2K)M^2 r+(4/3)(B^2+K)r^3 + M(6-6B^2 r^2
+4Kr^2) = 0.
\end{equation}
When $B^{2} \ll M$, and $K \ll M$, we have, that above some
critical region of values of $B$ and $K$, there are two null
circular orbits with radii

\begin{equation}
r_{1}= (3 M + 12M^3 K + (3M^3 + 120M^5 K)B^2,
\end{equation}
\begin{equation}
r_{2}=\frac{2 K(\sqrt{6} - 6 \sqrt{K}M)(1 + 2 K M^{2})
-B^{2}(\sqrt{6} +
\sqrt{K} M (-15 + 13 \sqrt{6 K} M + 6 K M^{2}))}{4 (K)^{3/2}}.
\end{equation}

\end{widetext}

When $B=K=0$, $r_{1}$ takes its Schwarzschild value $3 M$.
Unfortunately, we cannot find accurate values for the critical
region ($B_{cr}$, $K_{cr}$), because the values we get is quite
large, and, for instance for $K=B^{2}/2$ is about $0.189 M$, what
is on the boundary of applicability of the approximate metric
under consideration. Physical situation corresponds to some tidal
force $K$, which is larger than its pure Ernst value $B^{2}/2$.
Therefore, further we shall consider the new parameter $h$, which
is given by the relation:
$$ K = \frac{B^{2}}{2} + h.$$
Now, let us consider the effect of tidal gravitational attraction
of the surrounding structure upon such lens effects as light
bending angle and time delay. For this, let us follow the approach
of \cite{konoplya-lens1}.

If we know the distance of minimal approach $r_{min}$ with great
accuracy, we can perform integrations for finding bending angle:
$$ \alpha = \phi_{s}-\phi_{o} = $$
\begin{equation}
-\int_{r_{s}}^{r_{min}} \frac{d \phi}{d r} dr +
\int_{r_{min}}^{r_{o}} \frac{d \phi}{d r} dr - \pi.
\end{equation}

\begin{table}
\caption{Bending angle $\alpha$ and propagation time $\tau$ for
Ernst-Schwarzschild space-times (in geometrical units, $M=1$) for
$b=6$. "Observer" and "source"  are supposed to be situated not
far from the black hole in order to estimate influence of magnetic
field in the central region of the black hole: $r_{o} = r_{s}=
20$, $b=6$.}
\begin{tabular}{|c|c|c|c|}
  \hline
  $B$ & $h$ & $ \alpha + \pi$&  $t_{s} - t_{o} + d_{s-o}/\cos \mathcal{B}$ \\
  \hline
0& $ 0$  &                          4.252334 &  56.84554 \\
0& $ 5 \cdot 10^{-4}$ &             4.406231 &  57.25224 \\
0& $ 10^{-3}$&                      4.577176 &  57.72044  \\
$5 \cdot 10^{-4}$ & 0 &             4.252388 &  56.84567  \\
$5 \cdot 10^{-4}$&$5 \cdot 10^{-4}$&4.406293 & 57.25240 \\
$5 \cdot 10^{-4}$ & $10^{-3}$&      4.577247 & 57.72063 \\

\hline
\end{tabular}
\end{table}

Here $r_{o}$ is radial coordinate of an observer and  $r_{s}$ is
radial coordinate of the source.

In a similar fashion one can find the time delay, which is the
difference between the light travel time for the actual ray, and
the travel time for the ray the light would have taken in the
Minkowskian space-time:
\begin{equation}
t_{s}-t_{o} = -\int_{r_{s}}^{r_{min}} \frac{d t}{d r} dr +
\int_{r_{min}}^{r_{o}} \frac{d t}{d r} dr - \frac{d_{s-o}}{\cos \mathcal{B}} .
\end{equation}
Here the term  $\frac{d_{s-o}}{\cos \mathcal{B}}$ represents the
propagation time for a ray of light, if the black hole is absent.
The distance of minimal approach is the corresponding root of the
equation $dr/dt =0$. For pure Schwarzschild black hole it would be
the largest root, yet in our case it the second largest root of
the equation:
$$ r^3(-3+B^2(3 M^2 - 2 M r + r^2) +
          K(-6 M^2 + 4 M r + r^2)) $$
\begin{equation}
 - (2 M -
          r)(3 + 2 B^2 M r +
          K r (-4 M + 3 r) b^2 = 0.
\end{equation}
Looking numerically for the solution of the equation (30) one can
see that the tidal force $K$ pull out the radius of minimal
approach further from the black hole. We also can see from the
table I, that the presence of the mechanical structure leads to
increasing of the banding angle and time delay near the black
hole.

\section{Motion of massive particles}

The effective potential for massive neutral particles (21) is
shown in Figures 1 and 2 for zero and non-zero angular momentum
$L$ of the particle.

\begin{figure}
\resizebox{1\linewidth}{!}{\includegraphics*{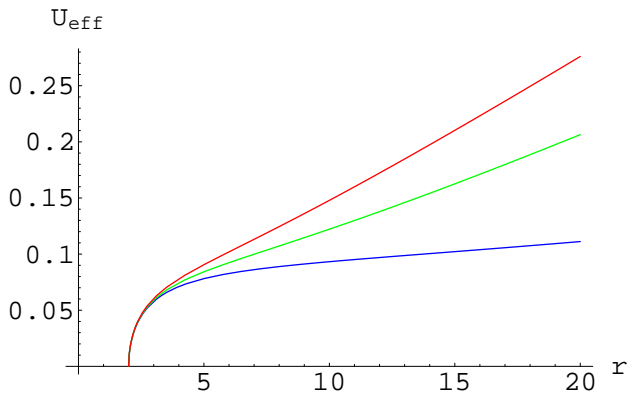}}
\caption{Effective potential for neutral particles:
$M=1$, $B=0.001$, $\mu =0.1$, $L=0$. $K=0.001$ (blue), $K=0.01$
(green), $K=0.02$ (red).}
\end{figure}

\begin{figure}
\resizebox{1\linewidth}{!}{\includegraphics*{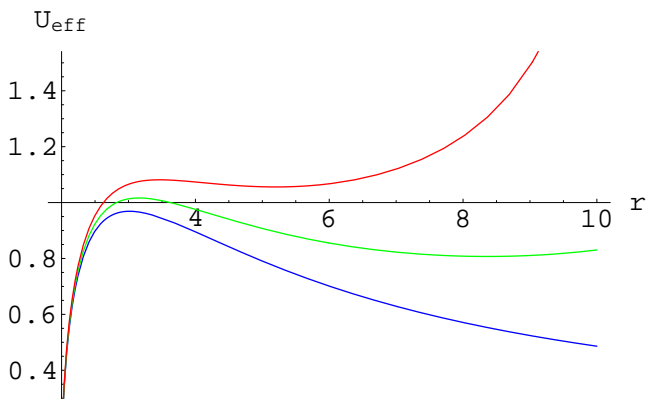}}
\caption{Effective potential for neutral particles:
$M=1$, $B=0.001$, $\mu =0.1$, $L=5$. $K=0.001$ (blue), $K=0.01$
(green), $K=0.02$ (red).}
\end{figure}

From the above figures one can see that the effective potential
can have form of the barrier or of a monotonically increasing
function.

For circular orbits the equation $V_{eff},_{r} = 0$ gives

\begin{equation}
-L^{2} (g_{tt},_{r} g_{\phi \phi} - g_{\phi \phi} g_{t t},_{r}) +
\mu^{2} g_{tt},_{r} g_{\phi \phi}^{2} =0.
\end{equation}

If one solves (27) for $L$, and uses it in the equation $dr/dt
=0$, or, equivalently, in
\begin{equation}
U_{eff}^{2} = E^{2},
\end{equation}
one obtains rather cumbersome system of equation for determination
of the parameters of orbits of massive particles.

The values $L/\mu$ and $E/\mu$ as a function of radius of circular
orbits are presented on Fig. 3 and 4 . They are found there from
accurate equations (31) and (32). From Fig. 3 and 4, one can see
that particle angular momentum and energy per units mass is
monotonically growing as functions of the radius of circular orbit
$r_{c}$, starting from some minimal value. This minimum value of
the test particle angular momentum corresponds to the orbit with
the innermost stable circular radius $r_{ic}$. The large $K$, the
more radius of the innermost stable circular orbit $r_{ic}$ is
pulled toward the black hole.

The binding energy is defined as the amount of energy that is
released by the test particle going from a stable circular orbit
$r_{c}$, to the innermost stable orbit of radius $r_{ic}$, i.e.

\begin{equation}
Binding \quad energy =
\frac{(E/\mu)_{r_{c}}-(E/\mu)_{r_{ic}}}{(E/\mu)_{r_{c}}}
\end{equation}

A test particle in an unstable circular orbits will fall into
black hole and the infall time is small compared to the radiative
time, so that the particle energy will be brought to the black
hole almost completely.

From the Figure (5) one can see the two features: first, that the
binding energy is greater for larger radius of circular orbit, and
this dependence on radius is strictly monotonic. Second, is that
the larger tidal force $K$, the larger the binding energy for a
given radius of the circular orbit of the particle $r_{c}$. The
last feature means that in presence of surrounding attracting
structure a test particle, when going from its stable orbit to the
innermost stable one, releases more energy, than it would release
without the above structure.

Finally, let us give the explicit form of the equations (31), (32)
through the orders $B^{2}$, $K$.

\begin{widetext}

\begin{equation}
\frac{E^{2}}{\mu^{2}} = \frac{1}{3} \left(\frac{r- 2 M}{r- 3 M}\right)^{2} \left(\frac{3}{r} (3 M - r) + 2 K (9 M^{2} - 10 M r + 3 r^{2}) - M(9 M - 4 r)B^{2}
\right)
\end{equation}

\begin{equation}
\frac{L^{2}}{\mu^{2}} = \frac{1}{3} \left(\frac{r}{r- 3
M}\right)^{2} (3 r (r- 3 M) - 18 M^{4} + 6 M^{3} r + 19 M^{2}
r^{2} - 14 M r^{3} + 3 r^{4} + M (9 M^{3} - 3 M^{2} r - 2 M r^{2}+
r^{3})B^{2}).
\end{equation}

\end{widetext}

These expressions are much simpler than exact equations (31), (32).

\begin{figure}
\resizebox{1\linewidth}{!}{\includegraphics{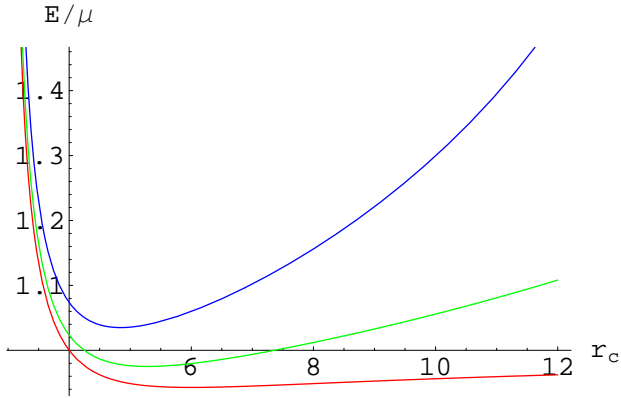}}
\caption{$E/\mu$ as a function of radius of the circular orbit $r_{c}$
$M=1$, $B=0.0001$, $K=(B^{2}/2)$ (bottom), $K=(B^{2}/2)+ 0.001$,
$K=(B^{2}/2) + 0.003$ (top).}
\end{figure}

\begin{figure}
\resizebox{1\linewidth}{!}{\includegraphics{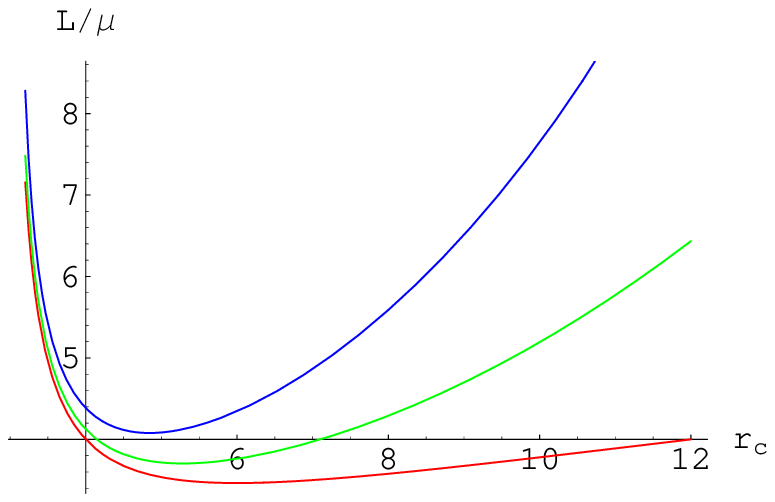}}
\caption{$L/\mu$ as a function of radius of the circular orbit $r_{c}$
$M=1$, $B=0.0001$, $K=(B^{2}/2)$ (bottom), $K=(B^{2}/2)+ 0.001$,
$K=(B^{2}/2)+ 0.003$ (top).}
\end{figure}

\begin{figure}
\resizebox{1\linewidth}{!}{\includegraphics{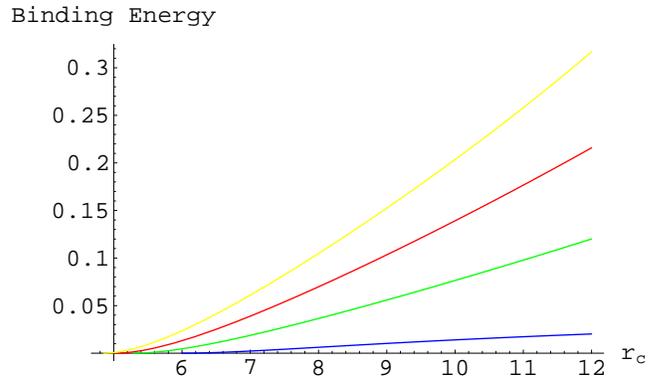}}
\caption{Binding energy as a function of radius of the circular orbit $r_{c}$
$M=1$, $B=0.0001$, $K=(B^{2}/2)$ (bottom), $K=(B^{2}/2)+ 0.001$,
$K=(B^{2}/2)+ 0.002$, $K=(B^{2}/2)+ 0.003$ (top).}
\end{figure}

\begin{figure}
\resizebox{1\linewidth}{!}{\includegraphics{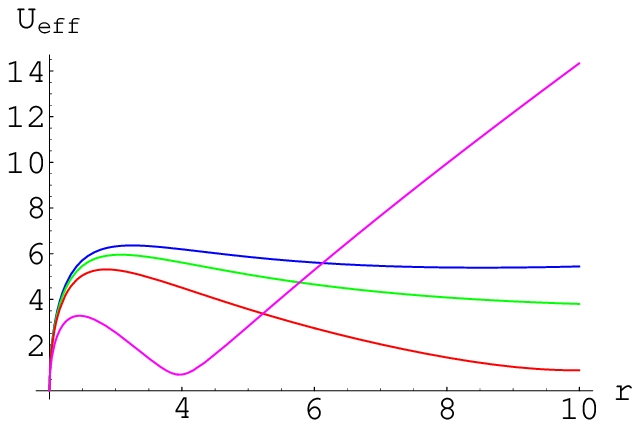}}0
\caption{Effective potential  for charged particles:  $e=0.6$ (top), $0$, $-0.6$, $-3.8$ (bottom),
$h=0$, $\mu M = 1$, $L=30$}
\end{figure}

\begin{figure}
\resizebox{1\linewidth}{!}{\includegraphics{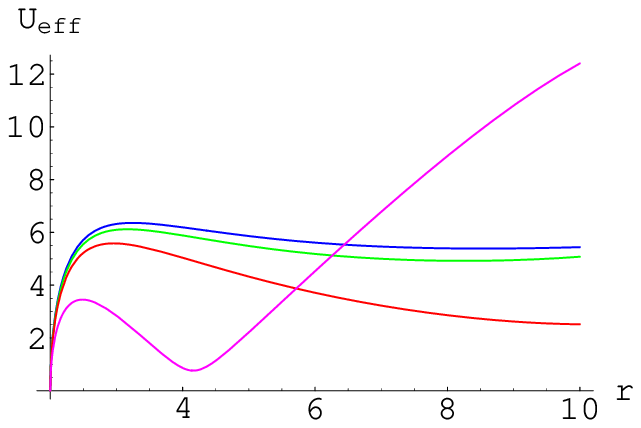}}
\caption{Effective potential  for charged particles:  $e=0.6$ (top), $0$, $-0.6$, $-3.8$ (bottom),
$h=0.005$, $\mu M = 1$, $L=30$}
\end{figure}

\begin{figure}
\resizebox{1\linewidth}{!}{\includegraphics{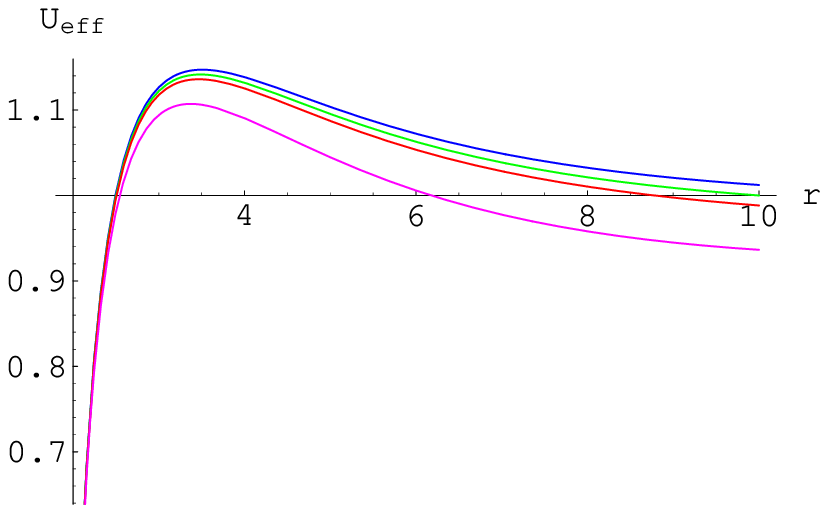}}
\caption{Effective potential for charged particles: $e B=0.006$ (top), $0$, $-0.006$, $-0.038$ (bottom),
$h=0$, $\mu M = 1$, $L=5$}
\end{figure}

\begin{figure}
\resizebox{1\linewidth}{!}{\includegraphics{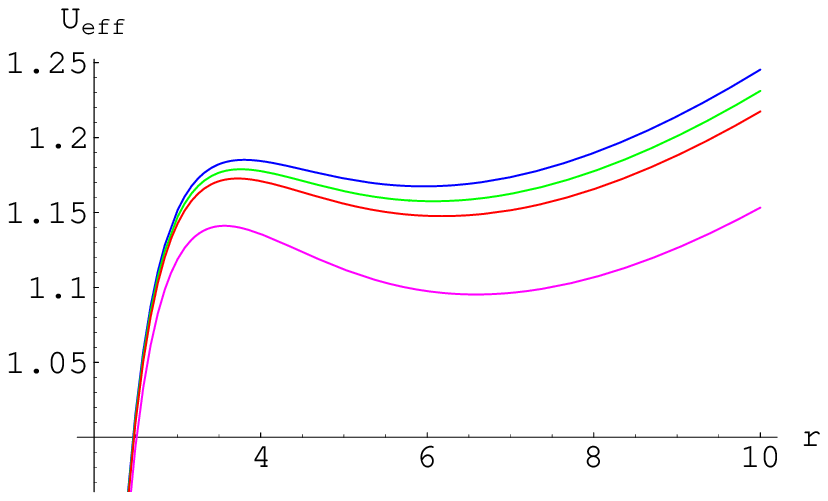}}
\caption{Effective potential for charged particles: $e B=0.006$ (top), $0$, $-0.006$, $-0.038$ (bottom),
$h=0.005$, $\mu M = 1$, $L=5$}
\end{figure}



For massive charged particles in vicinity of uncharged black
holes, one should change the angular momentum of the particle $L$,
to the generalized momentum:
\begin{equation}
L \rightarrow L + g_{\phi \phi} \frac{e B}{2}.
\end{equation}
The effective potentials for the case of charged particles are
shown on Fig. (6)-(8). The situation is dependent on the sign of
the charge, because the Lorentz force acting on the charged
particles has opposite directions for positive and negative
charges.

There are two reasons why we do not analyze the case of charged
particles in detail: First, the magnetic field used for
derivation of the considered metric is given only trough the first
order in $B$. Second, the effect of strong magnetic field for
charged particles is stipulated by the factor $e B/\mu$ (when
$M=1$), and is very large even for small $B \ll M$, because of the
large ratio $e/\mu$. Therefore, it is generally accepted, to
neglect "geometric" influence on propagation of charged particles,
and to consider the more realistic decaying magnetic fields on the
black hole background \cite{chargedpartciles}.

\section{Conclusion}

We have considered the motion of massless and massive test
particles near black holes immersed in asymptotically uniform
magnetic field and some surrounding structure which provides this
field. The tidal force from the surrounding structure has {\it
considerable} influence on the parameters of the test particle
motion. Let us enumerate them: a) it pulls radius of the circular
orbits off the black hole, b) increases the radius of minimal
approach for light, c) increases the time delay and bending angle
for light, d) increases the energy and momentum (per unit mass)
for a circular orbit of a given radius, e) increases the binding
energy of massive particles, which releases when a particle goes
from a given stable circular orbit to the innermost stable
circular orbit, f) the radius of the innermost stable circular
orbit is pulled closer to the black hole.

The used Preston-Poisson metric gives an excellent opportunity to
investigate the motion of test particles in the vicinity of
a supermassive "dirty" black hole, surrounded by some distribution
of matter and uniform magnetic field, and to approach, thereby, a
more realistic situation than that given by the Ernst solution.

\end{document}